\documentclass[fleqn]{article}
\usepackage{graphicx}
\date{}
\def\picture #1 by #2 (#3) {\vbox to #2 {
\hrule width #1 height 0pt depth 0pt
\vfill
\special{picture #3} } }
\usepackage{graphicx}
\begin{document}
\title{{\bf Quark-Gluon Coupling in the Global Colour Model of QCD}}
\vskip1.0cm
\author{{Reginald T. Cahill  and      Susan M. Gunner
  \thanks{E-mail: Reg.Cahill@flinders.edu.au,
Susan.Gunner@flinders.edu.au}}\\
{  }\\
  {Department of Physics, Flinders University}\\ { GPO Box 2100, Adelaide 5001,
Australia }\\
{November 1997}}

\maketitle

\begin{center}
\begin{minipage}{120mm}
\vskip 0.6in
\begin{center}{\bf Abstract}\end{center}
{The Global Colour Model of QCD is used in conjunction with
a pure-gluon lattice correlator (by Marenzoni {\it et al.}) to extract from meson data a 
momentum-dependent quark-gluon coupling down to $s \approx 0.3 GeV^2$ .This is compared
with a lattice calculation (by Skullerud) of the quark-gluon coupling.}\\


\end{minipage} \end{center}

\newpage
A key feature of Quantum Chromodynamics (QCD) is that the quark-gluon coupling varies
strongly with gluon momentum $q$ over the  range  $0<q<2GeV$ relevant to low energy
hadronic physics. Here we extract from meson data
this quark-gluon  coupling $g(q^2)$ down to $q=0.5 GeV$, using the Global Colour Model (GCM) of
QCD in conjunction with  a pure-gluon lattice correlator by Marenzoni {\it et  al.}
\cite{Marenzoni}. The extracted quark-gluon coupling (see Fig.2.) is compared with a recent
lattice calculation by Skullerud \cite{Skul}.

The GCM modelling of QCD is based on the idea that because  the hadronic correlators are given
by explicit functional integrals  it should be possible, after an appropriate change of
variables of integration, to identify a dominant configuration. It turns out that the existence
of this  dominant configuration is nothing more than the constituent quark effect.       The
GCM has been recently reviewed by Tandy
\cite{Tandy}, and applied to a number of problems \cite{TM, JFFhadt, FMchiral,
BFMformf,Meissner,FTua1,KTcond,sq}.

In the functional integral approach correlators are defined by 
\begin{equation} {\cal G}(..,x,...)=\frac{\int{\cal D}\overline{q}{\cal D}q{\cal
DA}{\cal D}\overline{C}{\cal
D}C....q(x).....\mbox{exp}(-S_{QCD}[A,\overline{q},q,\overline{C},C])} {\int{\cal
D}\overline{q}{\cal D}q{\cal DA}{\cal D}\overline{C}{\cal D}C
\mbox{exp}(-S_{QCD}[A,\overline{q},q,\overline{C},C])}.
\label{eq:2.1}\end{equation}
The  various complete correlators ${\cal G}$ lead to experimental observables. They 
are related by an infinite set of coupled Dyson-Schwinger Equations (DSE), and by the
Slavnov-Taylor gauge-symmetry-related identities (Non-scripted
$G$'s will denote constituent correlators, as defined later).  However the GCM does {\it not}
derive from these equations/identities, its nature follows instead from an analytical continuum
estimation  procedure for the functional integrations. Direct numerical estimation
procedures are  used in lattice modellings of the functional integrals.

The correlators in (\ref{eq:2.1}) may be extracted from the generating
functional of QCD 
\begin{equation}
Z_{QCD}[\overline{\eta},\eta,J]=\int {\cal D}\overline{q}{\cal D}q{\cal
D}A{\cal D}\overline{C}{\cal
D}C\mbox{exp}(-S_{QCD}[A,\overline{q},q,\overline{C},C]+\overline{\eta}q+
\overline{q}\eta+JA).
 \label{eq:2.2b}\end{equation}
The  functional
transformations which lead to the GCM are discussed in Tandy \cite{Tandy}; briefly  
and not showing source terms for convenience,  the gluon and ghost integrations are formally
performed
\begin{eqnarray*}\int {\cal D}\overline{q}{\cal D}q{\cal
D}A{\cal D}\overline{C}{\cal
D}C\mbox{exp}(-S_{QCD}[A,\overline{q},q,\overline{C},C])\end{eqnarray*}
\begin{eqnarray*} \mbox{\ \ \ \ \ \ \ \ \ \ \ \ }= \int {\cal D}\overline{q}{\cal D}q
\mbox{exp}(-\int 
\overline{q}(-\gamma . \partial+{\cal M})q +
\end{eqnarray*}
\begin{equation}\mbox{\ \ \ \ \ \ \ \ \ \ \ \ \ \ \ \ \ \ \ \ \ \ \ }
+\frac{g_0^2}{2}\int
j^a_{\mu}(x)j^a_{\nu}(y){\cal G}_{\mu\nu}(x-y)+\frac{g_0^3}{3!}\int
j^a_{\mu}j^b_{\nu}j^c_{\rho}{\cal
G}^{abc}_{\mu\nu\rho}+......)\label{eq:2.4}\end{equation}where
$j^a_{\mu}(x)=\overline{q}(x)\frac{\lambda^a}{2}\gamma_{\mu}q(x)$, $g_0$ is the bare coupling
constant, and ${\cal G}_{\mu\nu}(x)$ is  the  gluon
correlator  with no quark loops but including ghosts
\begin{equation}{\cal G}_{\mu\nu}(x-y)=
\frac{\int {\cal D}A{\cal D}\overline{C}{\cal
D}CA^a_{\mu}(x)A^a_{\nu}(y)\mbox{exp}(-S_{QCD}[A,\overline{C},C])}
{\int {\cal D}A{\cal D}\overline{C}{\cal
D}C\mbox{exp}(-S_{QCD}[A,\overline{C},C])}.
\label{eq:2.5}\end{equation}  
A variety of techniques for computing
${\cal G}_{\mu\nu}(x)$ exist:  the gluon-ghost DSE \cite{ANL}, and the gluon only
  DSE \cite{BP}  and lattice simulations \cite{Marenzoni,Lat}. The terms of higher order than
the term quartic in the quark fields   are difficult to explicitly  retain in any analysis. 
However we can model, in part, the effect of these higher order terms by replacing the
coupling constant
$g_0$ by a momentum dependent quark-gluon coupling $g(s)$, and neglecting terms like ${\cal
G}^{abc}_{\mu\nu\rho}$  and higher order. This $g(s)$ is a restricted form of vertex
function.  This modification 
$g_0^2{\cal G}_{\mu\nu}(p) \rightarrow D_{\mu\nu}(p) = g(p^2)^2{\cal G}_{\mu\nu}(p)$
and truncation in (\ref{eq:2.4}) then defines the GCM.  However we make one further
modification: we shall use lattice results with ghosts neglected for ${\cal G}_{\mu\nu}(x)$
\cite{Marenzoni}. Then $g(s)$ models as well the effect of the ghosts in both the gluon
correlator and the quark-gluon vertex. See \cite{ANL} for an analysis of these ghost effects.
We call $D_{\mu\nu}(p)$ the effective gluon correlator.

  The GCM
 is equivalent to using a quark-gluon field theory with the action
\begin{equation} S_{GCM}[A,\overline{q},q]=\int \left( 
\overline{q}(-\gamma . \partial+{\cal M}+iA^a_{\mu}\frac{\lambda^a}{2}\gamma_{\mu})q
 +\frac{1}{2} A^a_{\mu}D^{-1}_{\mu\nu}(i\partial)A^a_{\nu} \right).\label{eq:2.6}
\end{equation} 
Here $D_{\mu\nu}^{-1}(p)$ is the matrix inverse of $D_{\mu\nu}(p)$, which in turn is
the Fourier transform of $ D_{\mu\nu}(x)$. This action is invariant under $q\rightarrow Uq,
\overline{q}\rightarrow \overline{q}U^\dagger$, and $A^a_{\mu}\lambda^a \rightarrow U
A^a_{\mu}\lambda^a U^\dagger$  (where $U$ is a global $3\times3$ unitary colour matrix) - the
global colour symmetry of the GCM.
 The gluon  self-interactions  that
arise as a consequence of  the local colour symmetry in (\ref{eq:2.5}) and the ghost and vertex
effects   lead to 
$D_{\mu\nu}^{-1}(p)$ being  non-quadratic. Hence, in effect, the GCM models the QCD local
gluonic action  $\int F^a_{\mu\nu}[A]F^a_{\mu\nu}[A]$, having local colour symmetry, in
$S_{QCD}$ of (\ref{eq:2.1}), by a highly nonlocal action, having global colour symmetry, in the
last term of (\ref{eq:2.6}). The success of this modelling has been amply demonstrated
\cite{Tandy}. The form for
$g(p^2)$ is here determined by comparing  the meson data determined $D_{\mu\nu}(p)$  to the
pure-gluon lattice-determined correlator ${\cal G}_{\mu\nu}(p)$.

Hadronisation of the functional integrations in $(\ref{eq:2.1})$
 involves a sequence of  changes of
variables involving, in part, the transformation to bilocal boson fields, and
then to the usual local hadron fields (sources not shown): 
\begin{eqnarray*}  Z\approx\int {\cal D}\overline{q}{\cal D}q{\cal
D}A\mbox{exp}(-S_{GCM}[A,\overline{q},q]+\overline{\eta}q+
\overline{q}\eta)   \mbox{\ \ \ \ (GCM)}
\end{eqnarray*}  
\begin{equation}
\mbox{\ \ } =\int D{\cal B}...\mbox{exp}(-S[{\cal B},..]) \mbox{\ \ \ \
(bilocal fields)}
\label{eq:2.7}\end{equation} 
\begin{equation} \mbox{\ \ } =\int{\cal D}\pi{\cal
D}\rho{\cal D}\omega...\mbox{exp}(-S_{had}[\pi,\rho,\omega....]) 
  \mbox{\ \ \ \ (local fields) }.\label{eq:2.8}\end{equation}

The bilocal fields in (\ref{eq:2.7}) naturally arise and correspond to the fact that, for
instance, mesons are extended states.  This
bosonisation/hadronisation  arises   by functional integral calculus changes of
variables that are induced by generalized Fierz transformations that emerge from the colour,
spin and flavour structure of QCD \cite{RTC}.
The final functional integrations in (\ref{eq:2.8}) over the hadrons  give the
hadronic observables, and amounts to dressing each hadron by, mainly, lighter mesons.   
The basic insight is that the quark-gluon dynamics, in (\ref{eq:2.1}),  is
fluctuation dominated, whereas the hadronic functional integrations in  (\ref{eq:2.8}) are
not.  

The second key idea in the GCM is that in proceeding from  (\ref{eq:2.7}) to (\ref{eq:2.8})
one expands $S[{\cal B},..]$    about the configuration ${\cal
B}_{CQ}$ that  minimises it; giving  the GCM Constituent Quark (CQ) equations.
\begin{equation}
\frac{\delta S}{\delta {\cal B}(x,y)}\left|_{{\cal B}_{CQ}}=0\right.. 
\label{eq:Min}\end{equation}
 Thus for 
{\it all} hadrons one assumes a universal dominant configuration.  This amounts to assuming
that all hadrons share a  common dynamical feature.
Of the set ${\cal B}(x,y)_{CQ}$ only  $A(x-y)$ and $B(x-y)$ are
non-zero translation-invariant bilocal fields characterising the dominant configuration.
 Then writing out the translation invariant CQ equations
 we find that the dominant configuration is indeed simply the constituent  quark
effect as they may be written  in the form \cite{Tandy}, 
\begin{equation}
G^{-1}(p)=i\backslash \!\!\!p +m+
\frac{4}{3}\int\frac{d^4q}{(2\pi)^4}D_{\mu\nu}(p-q)\gamma_{\mu}G(q)\gamma_{\nu},
\label{eq:CQ}\end{equation}
and we see that this is  the gluon dressing of a constituent quark; and is  {\it exact}
in the GCM. Here 
\begin{equation}
G(q)=(iA(q)q.\gamma+B(q)+m)^{-1}=-iq.\gamma\sigma_v(q)+\sigma_s(q).
\label{eq:G}\end{equation}
 In
the chiral limit there are more ${\cal
B}_{CQ}$  fields that are non-zero, and a resultant degeneracy of the dominant
configuration is responsible for the masslessness of the pion \cite{Tandy}.  

 The  constituent quark  $G$ correlator should not be confused with the 
 complete quark correlator ${\cal G}$ from (\ref{eq:2.1}).    This ${\cal G}$ would be needed
to analyse the existence or otherwise of free quarks. The $G$ on the other hand relates
exclusively to the internal structure of hadrons, and to the fact that this appears to be
dominated by the constituent quark effect. The evaluation of  ${\cal G}$ is a very difficult
task, even in the GCM. $G$ is however reasonably easy to study using (\ref{eq:CQ}). 

The hadronic effective action in (\ref{eq:2.8}) arises when $S[{\cal B},..]$ is expanded
about the dominant CQ configuration: the 1st derivative is zero  by (\ref{eq:Min}), and the
2nd derivatives, or curvatures, give the constituent or core meson correlators $G(q,p;P)$
\begin{equation}
 G^{-1}(q,p;P)=F.T.\left(\frac{\delta^2 S}{\delta {\cal B}(x,y)\delta {\cal
B}(u,v)}\left|_{{\cal B}_{CQ}}\right.\right)
,
\label{eq:CM}\end{equation}
after exploiting the translation invariance and Fourier transforming. Higher order
derivatives  lead to couplings between the  meson cores.  The $G(q,p;P)$ are given by 
ladder-type  correlator equations, see \cite{Tandy}. The non-ladder effects can be inserted by
the final functional integrals in  (\ref{eq:2.8}), giving the complete GCM meson correlators
${\cal G}(q,p;P)$.   In the present analysis the $\omega$ and a$_1$ mesons are described by
these constituent meson correlators; that is, we ignore meson dressings of these mesons. 
The mass $M$ of these states is determined by finding the pole position of $G(q,p;P)$ in the
meson momentum $P^2=-M^2$, this leads to the homogeneous vertex equation
\begin{equation}\Gamma(p;P)=-\frac{4}{3}\int\frac{d^4q}{(2\pi)^4}D_{\mu\nu}(q-p)
\gamma_{\mu}G(q+\frac{P}{2})\Gamma(q;P)G(q-\frac{P}{2})\gamma_{\nu}\label{eq:meson}.
\end{equation}

 To solve $(\ref{eq:CQ})$
 for various 
$D_{\mu\nu}(p)$ and then to proceed to use $A(s)$ and $B(s)$ in  
  meson correlator equations for fitting observables to meson data is
particularly difficult. A  robust numerical technique is to use a separable
expansion for $D_{\mu\nu}(p-q)$ \cite{CG95,Burden}.  In  Landau gauge 
\begin{equation}
D_{\mu\nu}(p)=(\delta_{\mu\nu}-\frac{p_{\mu}p_{\nu}}{p^2})D(p^2), \mbox{\ \ and \ \ } 
{\cal G}_{\mu\nu}(p)=(\delta_{\mu\nu}-\frac{p_{\mu}p_{\nu}}{p^2}){\cal D}(p^2).
\label{eq:LG}\end{equation}
We expand   $D(p-q)$ in (\ref{eq:CQ}) into $O(4)$ hyperspherical harmonics
\begin{equation}
D(p-q)=D_0(p^2,q^2)+q.pD_1(p^2,q^2)+...
\end{equation}
where
\begin{equation}
D_0(p^2,q^2)=\frac{2}{\pi}\int_0^{\pi}d\beta \mbox{sin}^2 \beta D(p^2+q^2-2pq\mbox{cos}
\beta),...
\label{eq:D}\end{equation}
We then introduce multi-rank separable expansions for
each term
\begin{equation}
D_0(p^2,q^2)=\sum_{i=1,n} \Gamma_i(p^2)\Gamma_i(q^2),....
\label{eq:Gammas}\end{equation}
Introduction of the separable expansion clearly
breaks translational invariance and  must be regarded purely as a numerical
procedure,
much like a lattice breaks translation invariance. Translation invariance is
restored
as the rank of the separability is increased. Here we use  a rank $n=3$ form for $D_0$, and
rank $1$ form for $D_1$.
The constituent quark equations then have solutions of the form
\begin{equation}
B(s)=\sum B_i(s); \mbox{\ \  } B_i(s)=b_i\Gamma_i(s), ...
\label{eq:Bs}\end{equation}
 where the 
 $b_i,..$ are easily determined, in the chiral limit, to be 
\begin{equation}
 b_i^2=\frac{16}{3}\pi^2\int_0^{\infty} sds B_i(s)\sigma_s(s).
\label{eq:beq}\end{equation}
where
\begin{equation}
B_i(s)=\frac{\sigma_s(s)_i}{s\sigma_v(s)^2+\sigma_s(s)^2}
\label{eq:be}\end{equation}
and  $\sigma_s$ and $\sigma_v$ are  seen to have the form of sums
\begin{equation}
\sigma_s(s)=\sum_{i=1,n}\sigma_s(s)_i, \mbox{ \ \ \ \ }
\sigma_v(s)=\sum_{i=1,k}\sigma_v(s)_i ,
\label{eq:sigs}\end{equation}
However rather than specifying $\Gamma_i$ in (\ref{eq:Gammas}) we  proceed by
parametrising forms  for the $\sigma_s$ and $\sigma_v$; then the  $\Gamma_i$ follow from
(\ref{eq:Bs}) and (\ref{eq:be}): 
\begin{eqnarray*}
\sigma_s(s)_i=c_i\mbox{exp}(-d_is), i=1,2; \mbox{ \ \ \ \ }\sigma_s(s)_3=
c_3\left(\frac{2s-d_3(1-\exp(-2s/d_3))}{2s^2}\right)^2;\end{eqnarray*}
\begin{equation}
\sigma_v(s)=  \frac{2s-\beta^2(1-\exp(-2s/\beta^2))}{2s^2}.
\label{eq:seps}\end{equation}
As these forms are entire functions  we avoid spurious singularities developing
in $G$. The asymptotic
form of $\sigma_s(s)\sim 1/s^2$ for $s\rightarrow \infty$ is described by the  $\sigma_s(s)_3$
term.  With these parametrised forms  we can
numerically  relate the mass of the $a_1$ and $\omega$  mesons, from (\ref{eq:meson}) and
$f_{\pi}$ (for $N_f=2$) to the  chiral-limit parameter set $\{c_1,c_2,c_3,,d_1,d_2,d_3,\beta\}$
in a robust and stable manner.
 The  parameter values  are shown in Table 1. 
The chiral limit expression for $f_{\pi}$ is, see \cite{Tandy},
\begin{equation}
f_{\pi} =
6\int\frac{d^4q}{(2\pi)^4}\left(\sigma_v^2-2(\sigma_s\sigma_s'+s\sigma_v\sigma_v')
-s(\sigma_s\sigma_s''-(\sigma_s')^2)
-s^2(\sigma_v\sigma_v''-(\sigma_v')^2\right)B(q)^2.
\label{eq:fp}\end{equation}
 
The translation invariant form for the
effective gluon correlator is  easily reconstructed by using $D(p^2)=D_0(p^2,0)$ from
(\ref{eq:D})
\begin{equation}
D(p^2)=\sum_i\frac{1}{b_i^2}\frac{\sigma_s(0)_i}{\sigma_s(0)^2}\frac{\sigma_s(p^2)_i}
 {p^2\sigma_v(p^2)^2+\sigma_s(p^2)^2},
\end{equation}
 With the parameter set in Table 1, (\ref{eq:beq}) gives $b_1=0.0210~GeV^2$,
$b_2=0.0251~GeV^2$ and $b_3=0.0351~GeV^2$   and the resulting $D(p^2)$ is shown in Fig.1; 
it has estimated uncertainties of $5\%$.  
   Shown in Fig.1 for the pure gluon correlator is ${\cal D}(p^2)$ from the lattice
calculations, corresponding to the value
$\beta=6.0$, of Marenzoni {\it et al} \cite{Marenzoni} where the errors arise from a $5\%$
uncertainty in the lattice spacing; $a=0.50 \pm 0.025GeV^{-1}$. In Fig.2 we show the form of
$g(s)$, where 
$g^2(s)=D(s)/{\cal D}(s)$, see (\ref{eq:LG}), that then follows from our analysis.  Here the
error bars now indicate combined uncertainties.  This extracted quark-gluon coupling extends
down to
$0.3 GeV^2$,  and shows infrared (IR) enhancement. Below this limit the separable expansion
becomes unreliable unless  more terms and more fitting data are used. We have not corrected
for either lattice spacing dependence or for quark loops;  corrections for these would require
further development.
 It is possible to identify where the IR effect arises.  If we artificially lessen this
effect at small $s$ then we find that the main consequence is an increase in the value of
$f_{\pi}$. Indirectly,  then, we can show that the  IR signature is   the (inverse)  pion
size in comparison with the $a_1$ and
$\omega$ masses. The pion size enters through $f_{\pi}$  because in (\ref{eq:fp})  in the
chiral limit the pion form factor
$\Gamma_{\pi}(q;0)=B(q)$, see Tandy \cite{Tandy}.   However the GCM extraction  of this effect
does not explain what aspect of QCD drives it. In \cite{ANL} it is argued that
the QCD origin of this IR enhancement  is due to the ghost correlator  presence in  the
quark-gluon vertices.   We also report various condensate values that arise from the present work:
$<\overline{q}q>=(211.4\mbox{MeV})^3\mid_{1GeV}$,
$<g\overline{q}F_{\mu\nu}\sigma^{\mu\nu}q>=(491.5\mbox{MeV})^5\mid_{1GeV}$, and
$<\frac{\alpha}{\pi}F_{\mu\nu}F_{\mu\nu}>=0.026\mbox{GeV}^4\mid_{1GeV}$,  ignoring quark-loop
contributions.

Our most significant result follows from comparing, in Fig.2, the GCM-meson-data/lattice-gluon
determined quark-gluon coupling with that determined recently  by Skullerud \cite{Skul}
using a lattice calculation with $\beta=6.0(a=0.5\mbox{GeV}^{-1})$ and a lattice size of
$16^3\times48$. For comparison we also show the perturbative  coupling derived from the
two-loop beta function
\begin{equation}g^2(s)=\left(b_0\ln(\frac{s}{\Lambda^2})+\frac{b_1}{b_0}\ln\ln(\frac{s}{\Lambda^2}) 
\right)^{-1},\end{equation}
with $b_0=11/16\pi^2, b_1=102/(16\pi^2)^2$ for $\Lambda=0.420$GeV.  Fig.2 indicates a general
agreement of all three methods down to $s\approx0.7$GeV$^2$. The most significant difference
being the  decreasing lattice $g(s)$ in the deep IR; however this could be due to the finite
 lattice size which induces  an IR cutoff, or to the absence  of the ghost effects \cite{ANL}. 
The Skullerud data is similar to the running coupling extracted from the 3-gluon vertex
\cite{gvert}. In Fig.3 we  show   $\alpha_s=g^2/(4\pi)$
against $q (GeV)$. These
results indicate that QCD may now be sufficiently well modelled by the GCM in the low energy
regime that detailed  hadronic calculations may be performed, particularly for 
 the nucleon properties; the GCM having the advantage of easily dealing with the near chiral
limit needed for the nucleon, in contrast to lattice studies.  Fig.2 shows that the  lattice
results for the gluon correlator and the quark-gluon coupling may be combined to form
$D(s)_{lat}=g^2(s)_{lat}{\cal D}(s)_{lat}$; a lattice derived effective gluon correlator for
(\ref{eq:2.6}), except for the deep IR where we should be guided by the meson data fitting.  We
thus have a meeting of the continuum and lattice approaches. The deep IR behaviour remains
undetermined, but the region of uncertainty mainly affects questions of absolute confinement
and will have little effect upon low energy hadronic phenomena.

We thank N. Stella for assistance with the lattice results in \cite{Marenzoni}.
Research supported by an ARC 
Grant from Flinders University. This work is part
of the activities of the Special Research Centre for the Subatomic Structure of Matter,
University of Adelaide.

\vspace{5mm}

\begin{tabular}{llll}
\multicolumn{4}{l}  {Table 1: $\sigma_s(s)$ and $\sigma_v(s)$  Parameters}\\
\hline \hline

$\mbox{ \ \ }$$c_1$ &0.1732GeV$^{-1}$  & $d_1$    & 1.389GeV$^{-2}$\\
$\mbox{ \ \ }$$c_2$ &1.527GeV$^{-1}$  & $d_2$    & 4.982GeV$^{-2}$\\
$\mbox{ \ \ }$$c_3$  & 0.3435GeV$^{3}$ $\mbox{ \ \ \ \ \ \ \ \ \ \ \ \ \ \ \ \
\ }$ &$d_3$  & 1.971GeV$^{2}$
  \\
$\mbox{ \ \ }\beta$    & 0.4807GeV  &    \\
\hline

\end{tabular}
\newpage
\noindent {\bf Figure Captions}

\vspace{5mm}

\noindent  {\bf Figure 1} The  effective gluon correlator $D(s)$ (solid line)   extracted by
fitting the GCM to meson data.  Also shown are the lattice results for the pure gluon
correlator ${\cal D}(s)$ from Marenzoni {\it et al.} (1995). The error bars indicate 
 uncertainties  arising from the value of the lattice spacing $a=0.50 \pm 0.025GeV^{-1}$.
\vspace{5mm}

\noindent {\bf Figure 2} The  GCM quark-gluon coupling $g(s)$ (boxes). Here $g^2(s)$ was
obtained  by dividing the GCM effective gluon correlator $D(s)$ (solid line in Fig.1) by the
lattice gluon correlator ${\cal D}(s)$. The error bars arise from the lattice spacing
uncertainty and from  systematic errors in the fitting of the GCM to the meson data. Also
shown is $g(s)$ from the lattice calculation of Skullerud  (circles)  (1997). The curve shows
the two-loop form for $\Lambda=0.420$GeV.

\vspace{5mm}

\noindent {\bf Figure 3} Here we replot, for convenience, the GCM quark-gluon coupling in the
form
$\alpha_s=g^2/(4\pi)$ against $q (GeV)$.

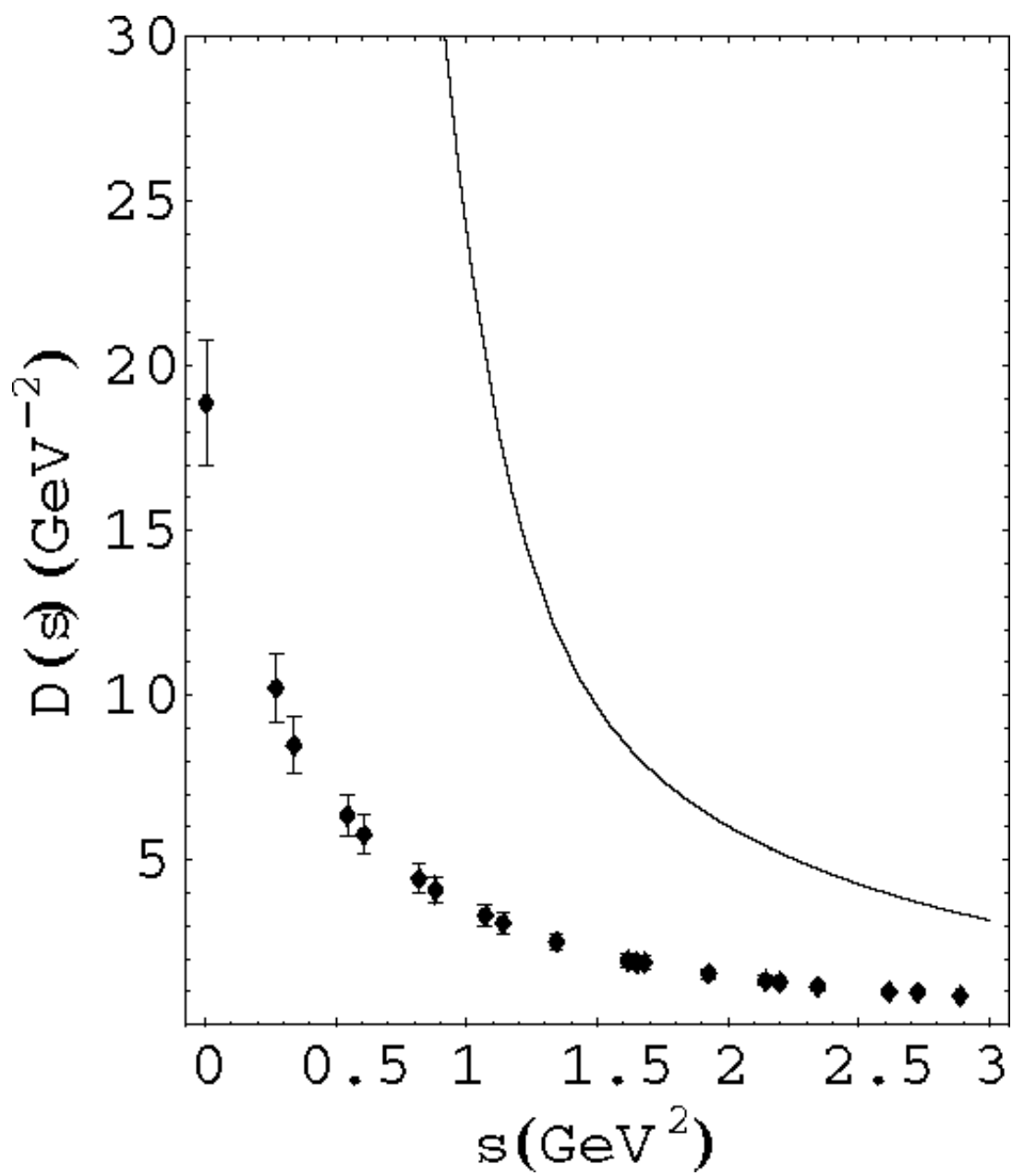
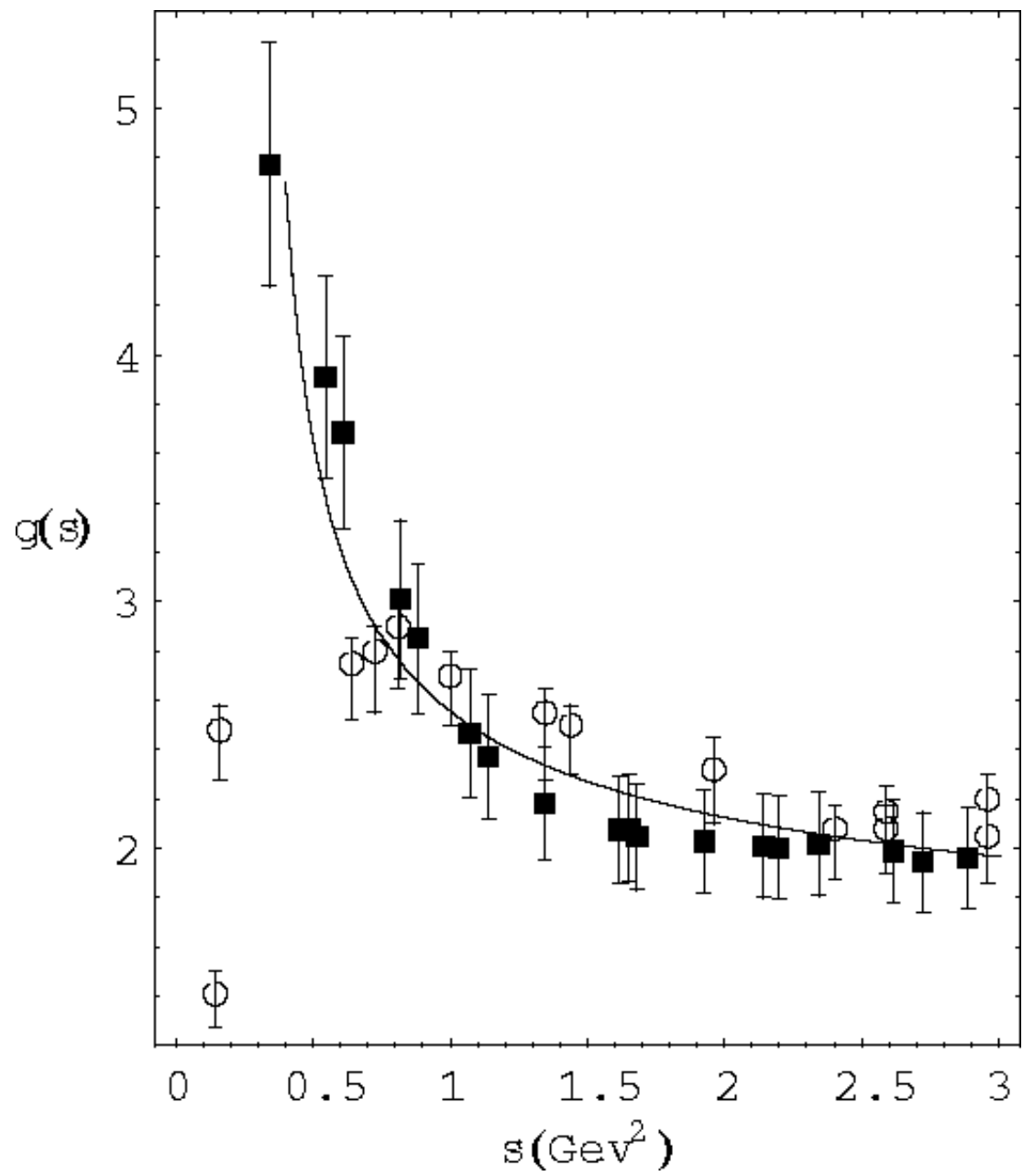
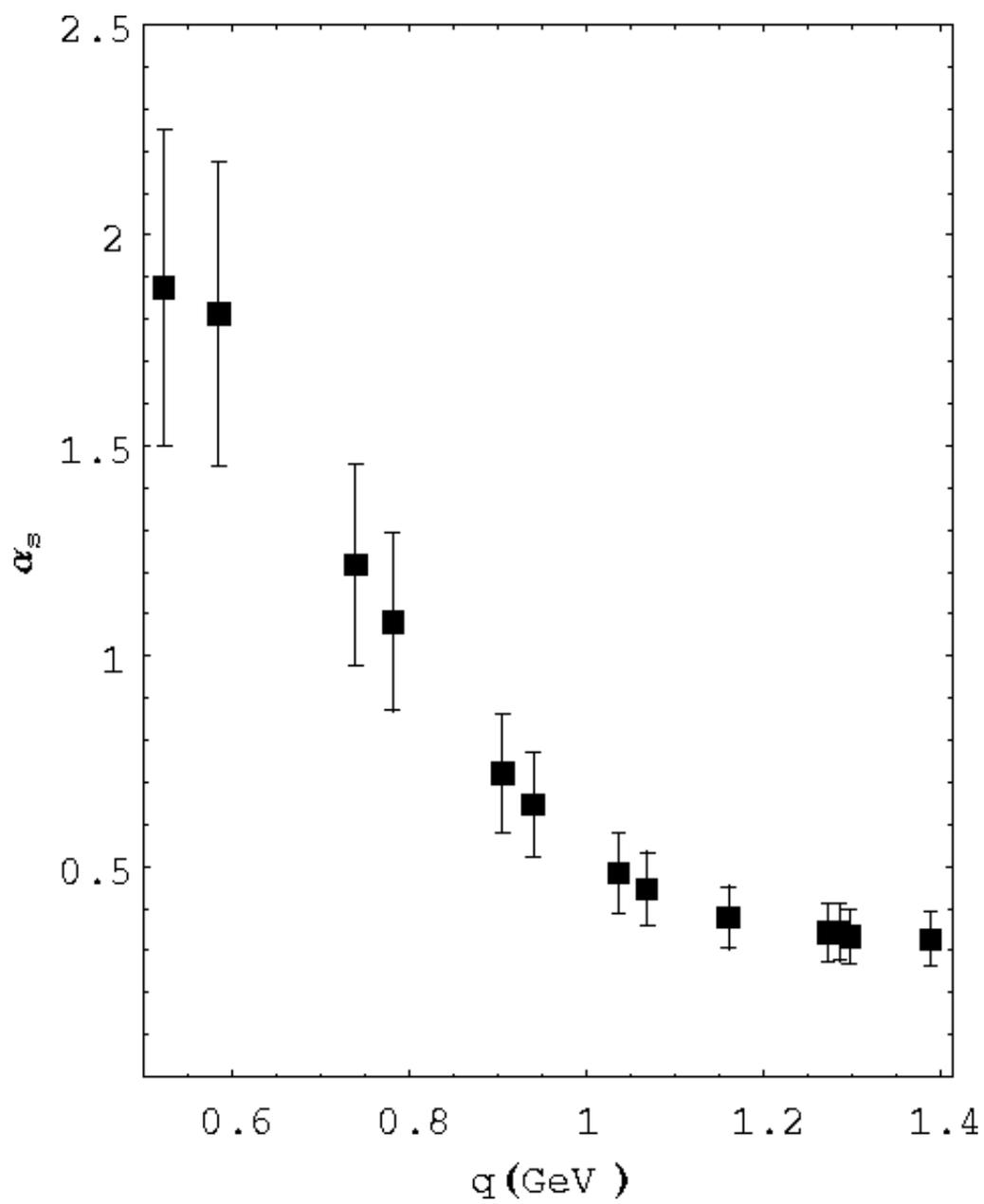
\end{document}